\documentclass[twocolumn,showpacs,amsmath,amssymb,aps,footinbib,prl,superscriptaddress]{revtex4-1}

\usepackage{graphicx}
\usepackage{dcolumn}
\usepackage{bm}
\usepackage{bbm}
\usepackage{hyperref}
\usepackage{xcolor}
\usepackage{accents}
\usepackage{dsfont}
\usepackage[normalem]{ulem}

\newcommand{\be}{\begin{equation}}
\newcommand{\ee}{\end{equation}}
\newcommand{\bea}{\begin{eqnarray}}
\newcommand{\eea}{\end{eqnarray}}

\usepackage{hyperref}

\usepackage{framed}

\begin{document}

\title{Tailoring the quantumness of a driven qubit}
\author{Alexander Friedenberger}
\affiliation{Department of Physics, Friedrich-Alexander-Universit\"at Erlangen-N\"urnberg, D-91058 Erlangen, Germany}

\author{Eric Lutz}
\affiliation{Department of Physics, Friedrich-Alexander-Universit\"at Erlangen-N\"urnberg, D-91058 Erlangen, Germany}
\affiliation{Institute for Theoretical Physics I, University of Stuttgart, D-70550 Stuttgart, Germany}

\begin{abstract}
We investigate the nonclassicality  of a  two-level system driven by an external time-dependent field in the presence of dephasing. We consider two criteria for nonclassicality, one based on  the quantum witness built upon  the no-signaling in time condition and one based on the temporal steering inequality, for a linearly driven qubit. We show  that both  strongly depend on the amplitude of the driving field. As a result,  we demonstrate that the quantumness of the two-level system can be successfully controlled by properly tuning the driving strength. We establish in particular that  maximal violations of classicality allowed by dephasing can be achieved at arbitrary chosen times.
\end{abstract}

\maketitle

Characterizing and controlling  nonclassicality   is of fundamental and practical importance. On the one hand, quantum features, such as coherence \cite{str17} and entanglement \cite{hor09}, play a crucial role in foundational tests of quantum theory \cite{leg02,rei09}. On the other hand, they have been identified as a  resource for  quantum technologies that  can  outperform their classical counterparts \cite{nie00}. Cases in point are quantum communication \cite{gis07}, quantum computation \cite{vin95} and quantum metrology \cite{gio11}. A common  method to establish  quantumness is to impose classical constraints that are violated by quantum mechanics. For instance, the classical assumptions of realism and locality lead to Bell's inequality \cite{bel64,bru14}, while those of macroscopic realism and noninvasive measurability result in the Leggett-Garg inequality \cite{leg85,ema14}.
 A violation of  the former reveals the presence of  nonclassical spatial correlations between two systems, whereas a  violation of the latter uncovers nonclassical temporal correlations in the dynamics of a single system. 
  A third, more recent approach is  based on the classical assumption of no-signaling in time, that is, the idea that a measurement does not affect the outcome  of a later measurement \cite{nor12,kof12}. Quantumness is here witnessed by the different population dynamics  of a system in the presence and  in the absence of a measurement. Experimental implementations of such quantum witness with single atoms \cite{rob15},   superconducting flux qubits \cite{kne16} and individual photons \cite{wan17,wan17a} have been reported lately.

 To our knowledge, nonclassicality as defined above has so far only been  investigated for time-independent qubits. However, most quantum systems of interest are driven by external fields. Prominent examples include laser driven atoms in quantum optics \cite{joa12}, driven spins in nuclear magnetic resonance \cite{van05},  driven quantum dots in condensed matter physics \cite{har05}, and laser driven chemical reactions \cite{bru92}, to name a few. External driving fields are known to offer new means to precisely control and manipulate the dynamics of quantum systems \cite{dit98}. They give rise, in addition,  to novel phenomena that are absent in stationary systems, such as stabilization against ionization \cite{rei70} and coherent destruction of tunneling \cite{gri98}. 
 
 In this paper, we perform a detailed analysis of the nonclassical properties of a  driven qubit with the help of the quantum witness based on the no-signaling in time condition \cite{nor12,kof12}. We show that the time evolution of the witness is strongly influenced by the external driving. In particular, while the position of the maxima of the  witness, which correspond to  maximal departure from classical behavior \cite{sch15}, are determined by the level splitting of the undriven qubit, they depend on  the amplitude  of the external linear field in the driven case.   We analytically determine the location of these maxima  for a two-level system subjected  to dephasing. We demonstrate as a result  that the maximum value of the quantum witness can be reached at an arbitrary chosen time by properly tuning the parameter of the driving field. We further apply these findings to temporal steering, a form of temporal correlations between the initial and final state of a quantum system \cite{che14,kar15,bar16,che16,bar16a,xio17}. The latter may be regarded as a temporal analog of the Einstein-Podolsky-Rosen  steering which allows to remotely prepare quantum states in another spatially-separated party \cite{ein35}. We specifically resolve the maxima of the  steering parameter for the driven qubit with dephasing and show that the latter may again be  controlled by the external driving field.

\textit{Driven qubit.}
We consider an externally  driven two-level system with Hamiltonian, $H(t)=\omega(t)\sigma_z/2$, where $\omega(t)$ is the time-dependent frequency and $\sigma_z$ the usual Pauli operator. For concreteness and simplicity, we shall assume a linear driving of the form, $\omega(t)=\omega_0+\Delta t$, with amplitude $\Delta$. The time evolution of a system operator $X$ may be described with the help of the Heisenberg equation of motion \cite{ali07},
\begin{equation}
\frac{dX}{dt}=i \left[H,X\right]+\frac{\partial X}{\partial t} + \frac{\gamma}{4}\left(\sigma_z[X,\sigma_z]+[\sigma_z,X]\sigma_z\right),
\label{eq:mastereq}
\end{equation}
where the last term on the right-hand side accounts for dephasing noise with intensity $\gamma$. This type of noise leads to phase damping in the eigenbasis of the qubit and is of importance in many applications \cite{nie00}.

\textit{Driven quantum witness.}
The quantum witness introduced in Refs.~\cite{nor12,kof12} exploits   the fact  that a quantum measurement of an observable necessarily disturbs any noncompatible observable. Consider an observable $A$ nonselectively measured at time $t=0$ and another observable $B$ measured at a later time $t>0$. The measurement outcome $n$ of $A$ is obtained with probability $p_n(0)$, while the measurement outcome $m$ of $B$ is obtained with probability $p'_m(t)$. For a joint measurement of the two observables, the probability for a $d$-level system of obtaining $m$ in the second measurement is,
\begin{equation}
p'_m(t)=\sum_{n=1}^d p(m,t |n,0)p_n(0),\label{eq:Prob}
\end{equation}
with the conditional probability $p(m,t |n,0)$.  In the absence of the first measurement on $A$, the probability of outcome $m$ of $B$ is denoted by $p_m(t)$. According to the classical no-signaling in time assumption, the nonselective measurement of  $A$ should have no influence on the statistical outcome of the later measurement of  $B$, and $p'_m(t) = p_m(t)$. The quantum witness is then defined as the difference $\mathcal{W}_q=| p_m(t)-p'_m(t)|$. A non-vanishing value of the witness is a clear signature of nonclassicality.

Specifying quantum behavior with the help of the quantum witness  presents a number of theoretical and experimental advantages: i) contrary to Bell's inequality, it applies to individual systems like the Leggett-Garg inequality, however,  ii) its implementation only requires two time measurements, in contrast to the three measurements usually needed to test the Leggett-Garg inequality, and iii) it   involves one-point expectations,  rather than two-point correlations, finally, iv) it provides a necessary and sufficient condition for macrorealism \cite{kof16}. In view of its simplicity, the quantum witness  is therefore ideally suited to investigate the influence of external driving.

In order to look for maximal violations of classicality for the driven qubit, we choose the noncommuting operators {$A= \sigma_x(\tau/2)$ and $B= \sigma_x(\tau)$}. We further initialize the system in the $|+\rangle$ eigenstate  of $\sigma_x$ \cite{sch15}.  We analyze the time evolution of the quantum witness by solving the Heisenberg equation of motion \eqref{eq:mastereq} for a set of basis operators consisting of $\{\sigma_x,\sigma_y, \sigma_z,I\}$ \cite{sup}. 
In the absence of the intermediate measurement, the probability to find the system at time $t=\tau$ in the state $|+\rangle$ reads \cite{sup},
\begin{eqnarray}
p_+(\tau)=\langle \Pi_{+}\rangle(\tau)=\frac{1}{2}\left[1+e^{-\gamma \tau}\cos\left(\frac{\Delta \tau^2}{2}+\omega_0 \tau\right)\right]
\end{eqnarray}
On the other hand, when a nonselective projective measurement of  $A= \sigma_x$ is performed at time $t=\tau/2$, the state of the system is updated to $\rho'=\sum_{i=\pm}\Pi_i \rho(\tau/2)\Pi_i$, where $\rho(\tau/2)$ denotes the density operator  prior to the measurement. The probability to find the system at time $t=\tau$ in the state $|+\rangle$ is now \cite{sup},
\begin{eqnarray}
{p}'_+(\tau)=\frac{1}{2}+\frac{e^{-\gamma \tau}}{4}\!\left[\cos\!\left(\frac{\Delta \tau^2}{4}\right)\!+\!\cos\!\left(\frac{\Delta \tau^2}{2}+\omega_0\tau \right)\right]
\end{eqnarray}
As a consequence, the linearly driven quantum witness $\mathcal{W}_q=|p_+(\tau)\rangle -p'_+ (\tau)|$ is given by the expression,
\begin{equation}
\label{5}
\mathcal{W}_q=\frac{e^{-\gamma \tau}}{4} \left| \cos \left(\frac{\Delta  \tau ^2}{4}\right)-\cos \left(\frac{\Delta  \tau ^2}{2}+\omega_0\tau \right)\right|.
\end{equation}
 The maximum value of the quantum witness \eqref{5} is determined by the number of possible outcomes of the intermediate nonselective measurement and is here given by $1/2$ \cite{sch15}. These upper bounds of quantumness are of special significance. A violation of the corresponding Tsirelson bound for Bell's inequality \cite{tsi80} would reveal new physics beyond quantum theory \cite{pop94,pop14} and have important consequences for quantum communication protocols \cite{bra06,fra11}. Furthermore, the temporal Tsirelson bound for the Leggett-Garg inequality \cite{fri10,bud13} has  been linked to the divisibility of quantum dynamics \cite{le17}.

 The maxima of the   quantum witness \eqref{5} for a given time $\tau$ are obtained by calculating the zeros of its derivative with respect of the driving amplitude $\Delta$. To simplify the discussion, we set $\omega_0=0$. We then find $(k \in \mathbb{Z})$ \cite{sup}, 
  \begin{align}
\Delta_1(k,\tau)&=\frac{4 \pi\left(2 k+1 \right)}{\tau ^2},\\
\Delta_2(k,\tau)&=\frac{4 \left(2 \pi  k-\tan ^{-1}\left(\sqrt{15}\right)\right)}{\tau ^2},\\
\Delta_3(k,\tau)&=\frac{4 \left(2 \pi  k+\tan ^{-1}\left(\sqrt{15}\right)\right)}{\tau ^2}.
\end{align}  
We note that all three equations (6)-(8) are independent of the dephasing constant $\gamma$.  

\begin{figure}[t]
\includegraphics[width=0.49\textwidth]{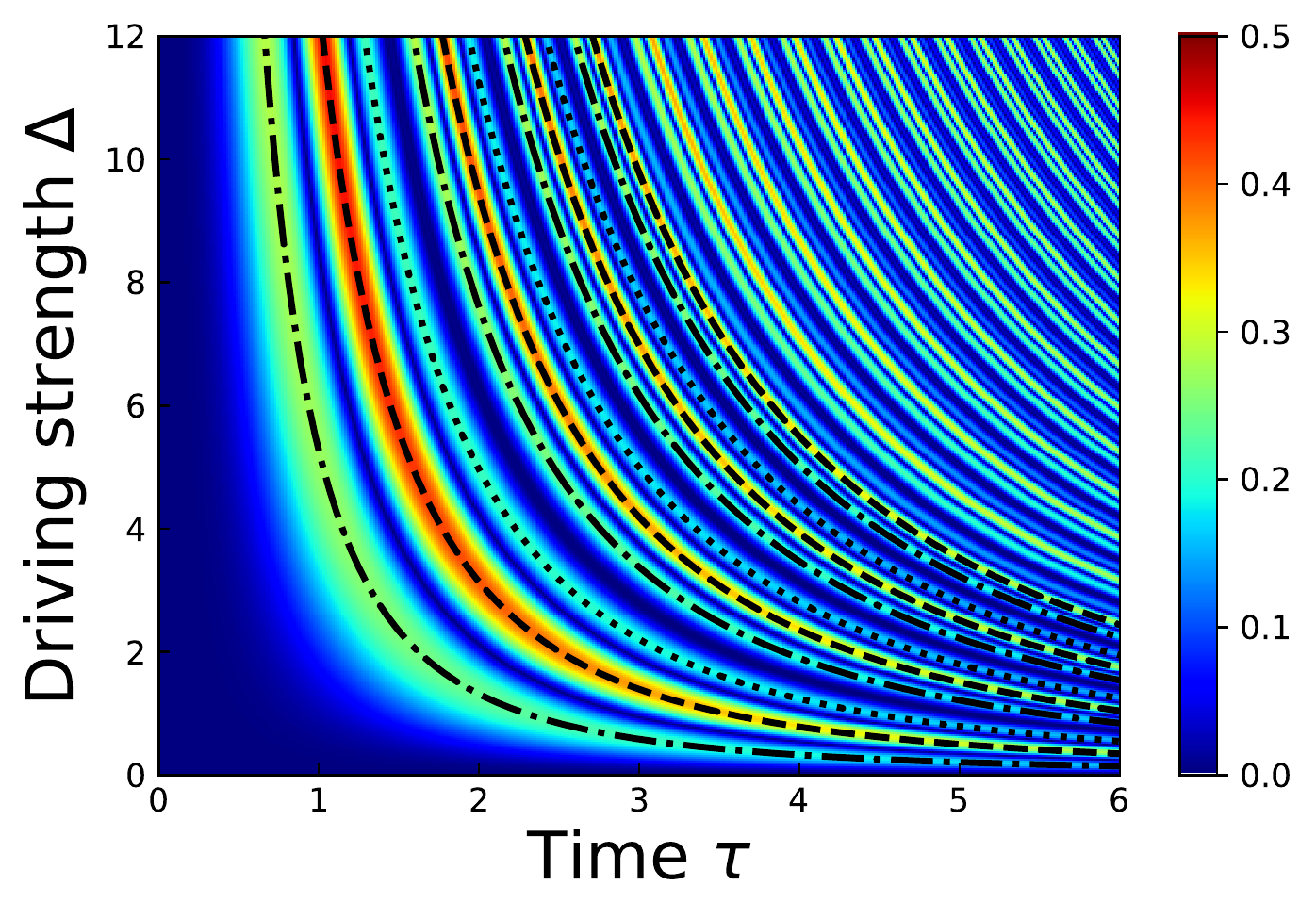}
\caption{Density plot of the quantum witness $\mathcal{W}_q$, Eq.~(5), as a function of time $\tau$ and driving amplitude $\Delta$, for a linearly driven two-level system with dephasing.  The black lines correspond to the maxima of the witness as determined by the respective conditions (6)-(8): $\Delta_1$ (dashed), $\Delta_2$ (dotted-dashed) and $\Delta_3$ (dotted). Parameters are $\omega_0=0$ and $\gamma = 0.1$. }
\label{fig:lin}
\end{figure}

 \begin{figure}[t]
\includegraphics[width=0.49\textwidth]{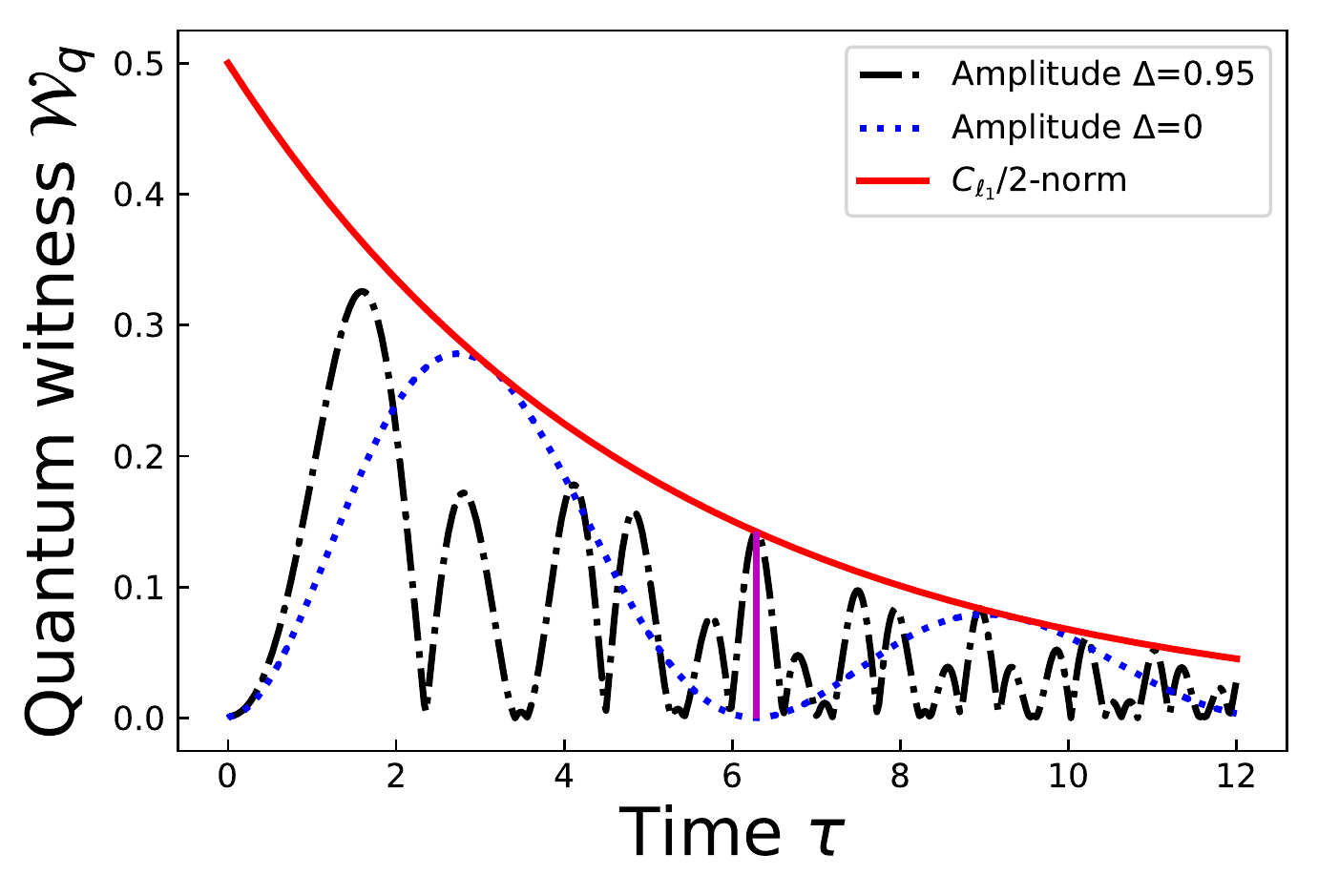}
\caption{Driven (black dotted-dashed) and undriven (blue dotted) quantum witness  $\mathcal{W}_q$, Eq.~(5), as a function of time $\tau$. While the maxima of $\mathcal{W}_q$, corresponding to maximal violations of classicality, are determined by the frequency $\omega_0$ of the undriven qubit, they are determined by both frequency and driving amplitude $\Delta$ for the linearly driven qubit. The minimum of the undriven  quantum witness at $\tau = 2\pi/\omega_0$ is turned into a maximum of the driven quantum witness for $\Delta = 0.95$ (vertical purple). The quantum witness is bounded from above by one-half the $l_1$-norm of coherence $C_{\ell1}$ (red solid). The dephasing rate is $\gamma=0.2$.}
\label{fig:lin}
\end{figure}

Figure 1 shows, as an illustration, the two-dimensional density plot of the quantum witness \eqref{5} as a function of the time $\tau$ and the driving strength $\Delta$. The maxima given by Eqs.~(6)-(8) are respectively represented by the black  dashed lines $(\Delta_1)$, the dotted-dashed lines $(\Delta_2)$ and the dotted lines $(\Delta_3)$. Figure 2 displays the time evolution of the quantum witness for various parameters.  The undriven ($\Delta= 0$) qubit  with dephasing ($\gamma =0.1$) (blue dotted)  exhibits oscillations with frequency $\omega_0=1$. It therefore vanishes for $\tau = n (2\pi /\omega_0), (n \in \mathbb{Z})$.  At these times, the nonclassicality of the undriven qubit cannot be established. An upper bound to the quantum witness is given by one half the $l_1$-norm, $C_{l_1}(\rho) = \sum_{i\neq j} |\rho_{ij}| = \exp(-\gamma \tau)$, (red solid) a coherence monotone which is  the sum of the modulus of the nondiagonal matrix elements of the density operator $\rho$ \cite{bau14,fri17}.
 By contrast, the behavior of the driven $(\Delta \neq 0$) qubit in the presence of dephasing ($\gamma =0.1$) (black dotted-dashed) is more involved. It presents  irregular oscillations whose maxima are still bounded by one half the $l_1$-norm of coherence. Their locations are determined by analogs to Eqs.~(6)-(8) for $\omega_0\neq 0$ which can be determined numerically. These criteria  can be successfully used to tailor the quantum witness such that it achieves its maximal possible value allowed by  dephasing at any  chosen time $\tau$. For example, the minimum of the witness of the undriven qubit at $\tau = 2\pi/\omega_0$ can be turned into a maximum by setting  the driving amplitude to $\Delta = 0.95$ as seen in the figure (vertical purple line). This point corresponds to the maximal possible violation of classicality at that time.
\begin{figure}[t]
\includegraphics[width=0.49\textwidth]{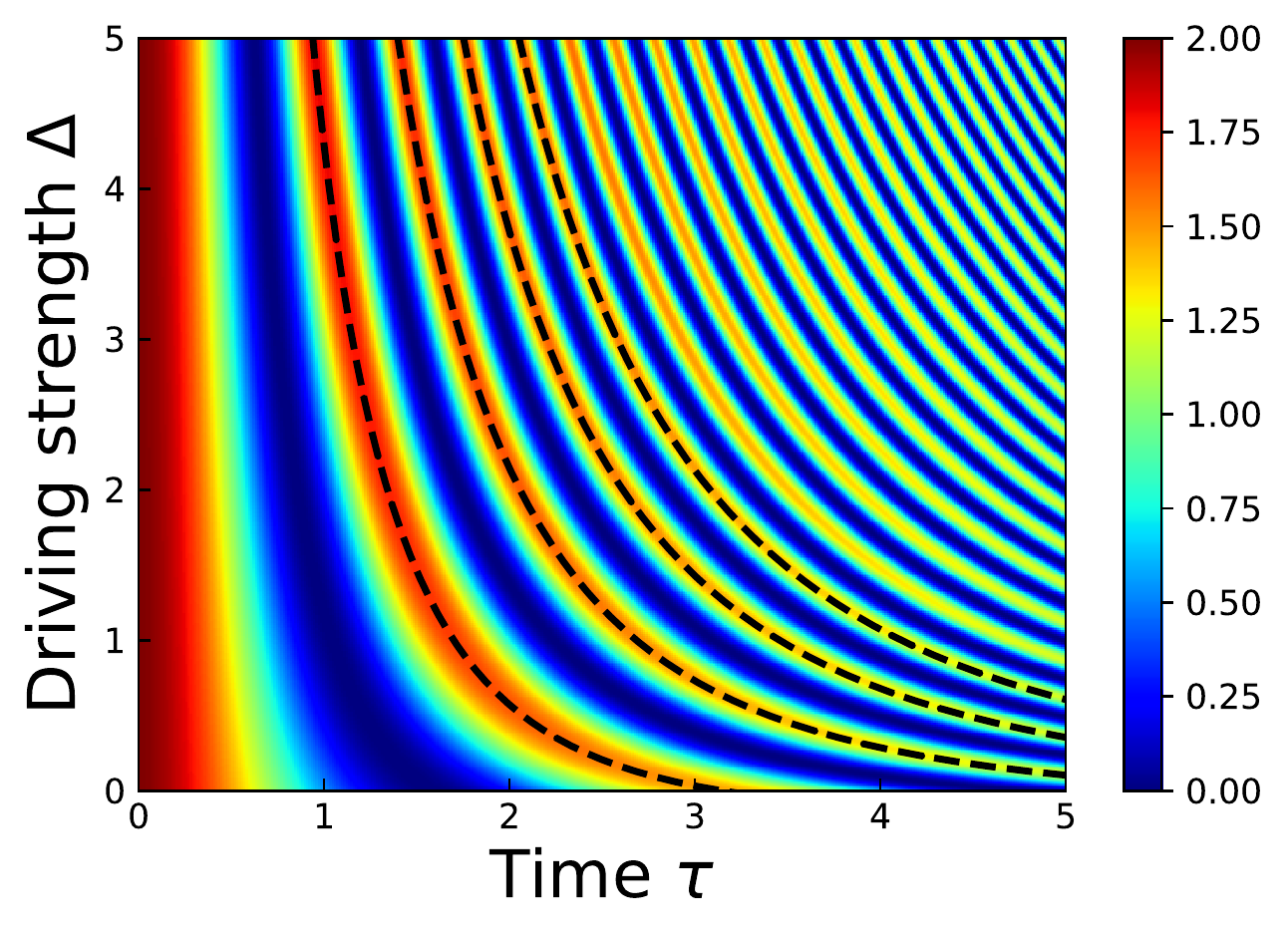}
\caption{Density plot of the steering function $S_2$, Eq.~(11), as a function of time $\tau$ and driving amplitude $\Delta$, for a linearly driven two-level system with dephasing.  The black dashed lines corresponds to the maxima of the steering parameter as determined by Eq.~(12). Parameters are $\omega_0=1$ and $\gamma = 0.1$. }
\label{fig:lin}
\end{figure}

\textit{Driven temporal steering inequality}.
We next apply these results to temporal steering. 
Temporal steering  is about the ability to affect the state of a  system at a later time through measurements, but  formulated in a two-party scenario \cite{che14,kar15,bar16,che16,bar16a,xio17}. It is    thus closely  related in spirit to the quantum witness.  Let us   consider two parties, Alice and Bob, that perform consecutive measurements (using a random sequence of mutually-unbiased bases known only to them) on one system to test temporal correlations between its initial and final states. Assuming that Alice measures observable $A$ at $t=0$ and Bob observable $B$ of the same qubit at time $t=\tau$, the following steering inequality holds for all classical states \cite{che14},
\begin{equation}
\label{8}
S_N(\tau)=\sum_{i=1}^N\mathbb{E}[\langle B_i\rangle^2_{A_i}]\le 1,
\end{equation}
where $N$ denotes the number of unbiased measurements performed by Bob. The expectation value,
\begin{equation}
\mathbb{E}[\langle B_i\rangle^2_{A_i}] = \sum_{a=\pm 1} P(a=A_i) \langle B_i\rangle^2_{A_{j}=a},
\end{equation}
is written in term of the  probability $ P(A_i=a) = \sum_\lambda q_\lambda P_\lambda(A_i=a)$ that depends on a classical variable $\lambda$ that specifies a given type of transmission channel with probability distribution $q_\lambda$. A violation of the steering inequality \eqref{8} indicates stronger than classical temporal correlations between initial and final states. Temporal steering has been successfully used to assess the security of quantum key distribution protocols \cite{kar15}. Violations of the steering inequality \eqref{8} have further been experimentally observed in photonic systems \cite{bar16a,xio17}.

\begin{figure}[t]
\includegraphics[width=0.49\textwidth]{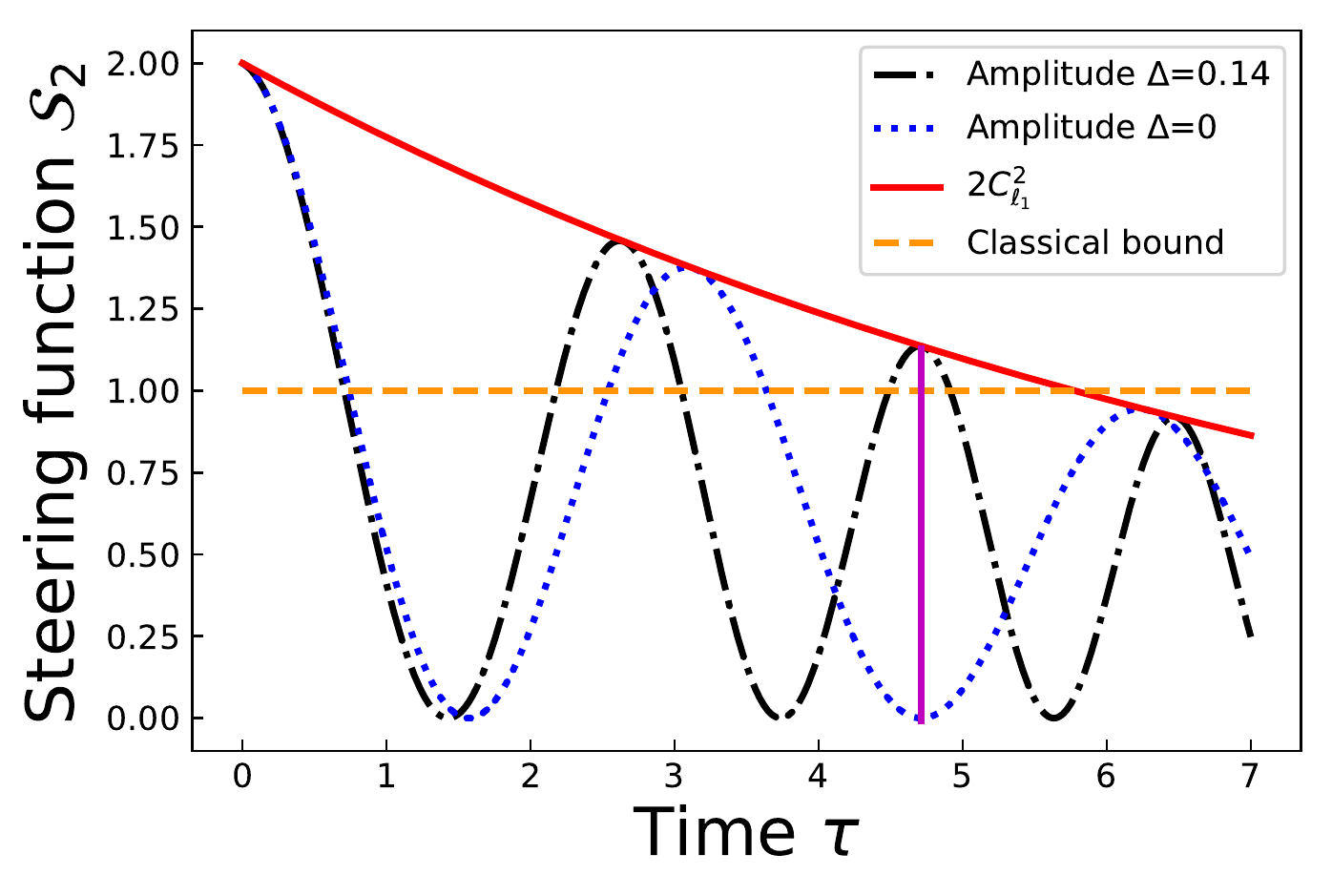}
\caption{Driven (black dotted-dashed) and undriven (blue dotted) steering parameter  $S_2$, Eq.~(11), as a function of time $\tau$. For short times, the steering inequality (9) is violated in both cases, indicating the presence of larger than classical temporal correlations. For long times, the inequality is always verified because of dephasing. Using the criterion (12), the position of the maxima of the driven steering function can be controlled via the driving amplitude $\Delta$. The minimum of the undriven $S_2$ at $\tau=3\pi/(2\omega_0)$ (no violation of the steering inequality) can thus be turned into  a maximum of the driven $S_2$ (violation of the steering inequality) by setting $\Delta=0.14$ (vertical purple). The steering parameter is bounded from above by twice the square of the $l_1$-norm of coherence $C_{\ell1}$ (red solid). Parameters are $\gamma=0.06$ and $\omega_0=1$.}
\label{fig:lin}
\end{figure}

We evaluate the steering parameter $S_2(\tau)$ for the linearly driven qubit with dephasing for $N=2$ by solving the Heisenberg equation (1) using  again the operator basis $\{\sigma_x,\sigma_y,\sigma_z,I\}$. We explicitly find \cite{sup},
\begin{equation}
S_2(\tau)=2 e^{-2\gamma \tau } \cos^2 \left(\frac{\Delta  \tau^2}{2}+\omega_0 \tau \right).\label{11}
\end{equation}
The maxima of the steering parameter  for a given time $\tau$ are determined by the zeros of the derivative of $S_2(\tau)$ with respect to the driving strength $\Delta$. We obtain \cite{sup},
\begin{equation}
\Delta(k,\tau)=\frac{2(k \pi -\omega_0\tau)}{\tau^2 },\quad k\in \mathbb{Z}.\label{12}
\end{equation}
The density plot of the steering function $S_2(\tau)$ as a function of the time $\tau$ and the driving parameter $\Delta$ is shown in Fig.~3 together with  the maxima specified by Eq.~\eqref{12} (black dashed). The undriven $(\Delta =0)$ steering parameter  with dephasing $(\gamma = 0.06)$ is represented as a function of time in Fig.~4. We observe that  $S_2(\tau)$ exhibits oscillations with frequency $\omega_0=1$ that are exponentially damped because of  dephasing. The upper bound to the steering function is given by $2 C_{l_1}^2 = 2 \exp(-2\gamma \tau)$. The steering inequality (9) is always violated in the limit $\tau\rightarrow 0$, demonstrating the presence of stronger  than classical temporal correlations, whereas  it is always verified in the limit $\tau\rightarrow \infty$. At intermediate times, the steering inequality is sometimes violated, sometimes not, owing to the oscillatory behavior of $S_2(\tau)$. The presence of stronger than classical temporal correlation can therefore not always be definitely established at these times. The driven steering parameter \eqref{11} displays a  similar behavior (black dotted-dashed). However, in the driven case the positions of its maxima are determined by both the frequency $\omega_0$ and  the amplitude $\Delta$ as given by Eq.~\eqref{12}. Similarly to the driven witness (5), the driving amplitude $\Delta$ can be tuned so that the steering function $S_2(\tau)$ reaches its maximum value permitted by dephasing at any arbitrary time $\tau$. For instance, the minimum of the undriven steering parameter at $\tau=3\pi/(2\omega_0)$, which corresponds to no violation of the steering inequality, can be successfully turned into a maximum of the driven steering function for $\Delta = 0.14$, resulting in a violation of Eq.~(9).

\textit{Conclusions.}
We have investigated the quantumness of a linearly driven two-level system based on the quantum witness as well as the temporal steering inequality. We have in both cases analytically determined the positions of maximal departures from classicality which strongly depend on the driving amplitude. We have demonstrated that the latter can be exploited  to tailor the nonclassicality of the qubit. Concretely, a minimum of either the undriven quantum witness or the undriven steering function, that exhibits no sign of quantumness, can be turned into a maximum of the respective driven quantity, at an arbitrary chosen time,  by simply tuning the driving amplitude. As a consequence, maximal nonclassicality, as permitted by dephasing, may be achieved even with minimal experimental control on the system.

\begin{appendix}
\section{Supplemental Material}
\subsection{Calculating the quantum witness}
The Heisenberg equation of motion \eqref{eq:mastereq} may be expressed in the operator basis $\{\sigma_x,\sigma_y,\sigma_z,I\}$ as,
\begin{equation}
\frac{d}{dt}{\begin{pmatrix}\sigma_x\\ \sigma_y\\ \sigma_z\\ I\end{pmatrix}}=\left(
\begin{array}{cccc}
 -\gamma  & -\omega (t) & 0 & 0 \\
 \omega (t) & -\gamma  & 0 & 0 \\
 0 & 0 & 0 & 0 \\
 0 & 0 & 0 & 0 \\
\end{array}
\right).\label{eq:meqOp}
\end{equation}
The time evolution of an operator $X$ is given by $X_H(t) = V^\dagger(0,t) X$ \cite{ali07}.
Restricting the analysis to $\{\sigma_x,\sigma_y\}$, the propagator $V^\dagger(t_1,t_2)$ follows from Eq.~\eqref{eq:meqOp} as,
\begin{widetext}
\begin{equation}
V^\dagger(t_1,t_2)=\left(
\begin{array}{cc}
 e^{-\gamma  \left(t_2-t_1\right)} \cos \left[\frac{1}{2} \left(t_1-t_2\right) \left(\Delta  \left(t_1+t_2\right)+2 \omega _0\right)\right] & e^{-\gamma  \left(t_2-t_1\right)} \sin \left[\frac{1}{2} \left(t_1-t_2\right) \left(\Delta  \left(t_1+t_2\right)+2 \omega _0\right)\right] \\
 -e^{-\gamma  \left(t_2-t_1\right)} \sin \left[\frac{1}{2} \left(t_1-t_2\right) \left(\Delta  \left(t_1+t_2\right)+2 \omega _0\right)\right] & e^{-\gamma  \left(t_2-t_1\right)} \cos \left[\frac{1}{2} \left(t_1-t_2\right) \left(\Delta  \left(t_1+t_2\right)+2 \omega _0\right)\right] \\
\end{array}
\right).\label{eq:prop}
\end{equation}
\end{widetext}
We further need to  account for the complete dephasing induced by  the nonselective projective measurement. During a nonselective measurement of $\sigma_x$, the expectation value of $\sigma_x$ remains unchanged. However, the expectation values of $\sigma_z$ and $\sigma_y$ will be set to zero. We describe that operation with the matrix,
\begin{equation}
\delta_x=\begin{pmatrix}1&0\\0&0\end{pmatrix},
\end{equation}
in the same basis that we have used for the propagator. The probability to find the system in state $|+\rangle$ is given by the expectation value of the projector $\Pi_+={(I+\sigma_x)}/{2}$.
Assuming the system to be initially in state $|+\rangle$, we find in absence of the  intermediate nonselective measurement,
\begin{equation}
\langle \sigma_x\rangle (\tau)=V^\dagger(0,\tau)\cdot (1,0)^T= e^{-\gamma  \tau } \cos \left(\frac{\Delta  \tau ^2}{2}+  \omega _0 \tau\right).
\end{equation}
As a result, the probability to find the system in state $|+\rangle$ at time $\tau$ is,
\begin{equation}
p_+(\tau)=\frac{1}{2}\left[1+e^{-\gamma  \tau } \cos \left(\frac{\Delta  \tau ^2}{2}+  \omega _0 \tau \right)\right]. 
\end{equation}
On the other hand, in the presence of a nonselective $\sigma_x$ measurement  at time $\tau/2$, we obtain,
\begin{equation}
\begin{aligned}
\langle \sigma_x\rangle (\tau)&=V^{\dagger} (\tau/2,\tau)\cdot\delta_x \cdot V^\dagger(0,\tau/2)\cdot (1,0)^T\\
&=\frac{1}{2} e^{-\gamma  \tau } \left[\cos \left(\frac{\Delta  \tau ^2}{2}+\tau  \omega _0\right)+\cos \left(\frac{\Delta  \tau ^2}{4}\right)\right].
\end{aligned}
\end{equation}
The probability to find the system in state $|+\rangle$ at time $\tau$ is now, 
\begin{equation}
p_+'(\tau)=\frac{1}{2}+\frac{e^{-\gamma \tau}}{4}\left[\cos\left(\frac{\Delta \tau^2}{4}\right)+\cos \left(\frac{\Delta  \tau ^2}{2}+  \omega _0 \tau \right)\right].
\end{equation}
The quantum witness follows as $\mathcal{W}_q=|p_+(\tau)\rangle -p'_+ (\tau)|$.

\section{Maxima of the quantum witness}
In order to determine the maxima of the quantum witness \eqref{5} for a given  time $\tau$, we compute its partial derivative with respect to the driving amplitude $\Delta$. We obtain,
\begin{widetext}
\begin{equation} 
\frac{\partial }{\partial \Delta}\mathcal{W}_q(\tau)=-\frac{1}{16} \tau ^2 e^{-\gamma  \tau } \left[\sin \left(\frac{\Delta  \tau ^2}{4}\right)-2 \sin \left(\frac{2\Delta  \tau ^2+4\omega_0\tau}{4}\right)\right] \text{sgn}\left[\cos \left(\frac{\Delta  \tau ^2}{4}\right)-\cos \left(\frac{2\Delta  \tau ^2+4\omega_0\tau}{4}\right)\right],
\end{equation}
\end{widetext}
where $\text{sgn}(x)$ is the sign function. The sign function only vanishes when  $\mathcal{W}_q=0$, that is, for minima of the quantum witness. Similarly, the prefactor only vanishes for $\tau=0$, which is again  a minimum of the witness. The driving amplitude thus satisfies the equation,
\begin{equation}\label{19}
\sin \left(\frac{\Delta  \tau ^2}{4}\right)-2 \sin \left(\frac{2\Delta  \tau ^2+4\omega_0\tau}{4}\right) {=}0.
\end{equation} 
To simplify the discussion, we consider the case $\omega_0=0$. Equation \eqref{19} has then four solutions. The first solution,
  \begin{equation}
\Delta_{0}(k,\tau)=\frac{8 \pi  k}{\tau ^2}, \quad \,k \in \mathbb{Z}
\end{equation}
 corresponds to the situation where both sine functions  vanish, that is, to a minima of the quantum witness.  The maxima of the quantum witness are finally determined by the remaining three solutions,
 \begin{align}
\Delta_1(k,\tau)&=\frac{4 \pi\left(2 k+1 \right)}{\tau ^2},\\
\Delta_2(k,\tau)&=\frac{4 \left(2 \pi  k-\tan ^{-1}\left(\sqrt{15}\right)\right)}{\tau ^2},\\
\Delta_3(k,\tau)&=\frac{4 \left(2 \pi  k+\tan ^{-1}\left(\sqrt{15}\right)\right)}{\tau ^2}.
\end{align}     
The solution  $\Delta_1(k,\tau)$ saturates the dimensional bound of $1/2$ of the quantum witness in the absence of dephasing. The positions of the maxima for $\omega_0\neq 0$ may be found numerically.

\section{Calculating the steering parameter}
In the temporal steering scenario, Alice starts by measuring one of a set of $N$ observables, before handing the system over to Bob, who in turn measures the same observable as Alice \cite{che14,kar15,bar16,che16,bar16a,xio17}. Independent of the initial state of the system, the measurement performed by Alice projects the state into one of the eigenstates of the measured observable. In the current scenario, the measured observables are the two Pauli operators $\sigma_x$ and $\sigma_y$. Assuming that the initial state is a maximally mixed state $\rho=\mathds{1}/2$, the measurement will yield either of the two possible measurement outcomes with equal probability. We thus consider the time evolution of the system for the four possible post-measurement states after the initial measurement by Alice, which correspond to  the $\sigma_x$ eigenstates  $|+\rangle,|-\rangle$ and the $\sigma_y$ eigenstates $|\phi^+\rangle,|\phi^-\rangle$, respectively.
Using the propagator \eqref{eq:prop} with,
\begin{equation}
|\pm\rangle{=}\begin{pmatrix}\pm1\\0\end{pmatrix}; \quad |\phi^{\pm}\rangle {=} \begin{pmatrix}0\\ \pm 1\end{pmatrix}.
\end{equation}
we find the same contribution in all four cases,
 \begin{equation}
\langle \sigma_{x,y}\rangle^2(\tau)=e^{-2\gamma \tau } \cos^2 \left(\frac{\Delta  \tau^2}{2}+\omega_0 \tau\right),
\end{equation}
which directly leads to the steering parameter  \eqref{11}.

\section{Maxima of the steering parameter}
To determine the maxima of the steering parameter \eqref{11} for a given time $\tau$,  we evaluate its partial derivative with respect to the driving amplitude $\Delta$,
\begin{equation}
\frac{\partial }{\partial \Delta} S_2(\tau)=-\tau^2 e^{-2 \gamma  \tau} \sin  (\Delta  \tau^2+2 \omega_0\tau).
\end{equation}
The value $\tau=0$ corresponds to a trivial maximum. We are thus left with the equation,
$\sin(\Delta \tau^2+2\omega_0\tau){=}0$,
which is fulfilled for $\Delta \tau^2+2\omega_0\tau {=} k\pi ,( k \in \mathbb{Z})$, or,
\begin{align}
\Delta_0(k,\tau)&=\frac{ (2 k+1) \pi -2\omega_0\tau}{\tau^2 }\\
\Delta_1(k,\tau)&=\frac{ 2 k \pi -2\omega_0\tau}{\tau^2 }.
\end{align}
Here, $\Delta_0(k,\tau)$ refer to minima, while $\Delta_1(k,\tau)$ refer to maxima. 
\end{appendix}

\begin{thebibliography}{99}
\bibitem{com} It is easy to see that the following contradiction $4j = 2k + 1$ is obtained when the first cosine is equal to $-1$ and the second cosine to $1$.
\end{thebibliography}
\end{document}